\documentclass[preprint]{aastex631}
\usepackage{gensymb}
\usepackage{color}
\usepackage{lineno}

\shortauthors{Zhang et al.}
\begin{document}
\title{Birth places of extreme ultraviolet waves driven by impingement of solar jets upon coronal loops}

\author{Liang Zhang}
\affiliation{Shandong Key Laboratory of Optical Astronomy and Solar-Terrestrial Environment, School of Space Science and Physics, Institute of Space Sciences, Shandong University, Weihai, Shandong, 264209, China; ruishengzheng@sdu.edu.cn}
\affiliation{CAS Key Laboratory of Solar Activity, National Astronomical Observatories, Chinese Academy of Sciences, Beijing 100101, People's Republic of China}

\author{Ruisheng Zheng}
\affiliation{Shandong Key Laboratory of Optical Astronomy and Solar-Terrestrial Environment, School of Space Science and Physics, Institute of Space Sciences, Shandong University, Weihai, Shandong, 264209, China; ruishengzheng@sdu.edu.cn}
\affiliation{CAS Key Laboratory of Solar Activity, National Astronomical Observatories, Chinese Academy of Sciences, Beijing 100101, People's Republic of China}

\author{Huadong Chen}
\affiliation{CAS Key Laboratory of Solar Activity, National Astronomical Observatories, Chinese Academy of Sciences, Beijing 100101, People's Republic of China}

\author{Yao Chen}
\affiliation{Shandong Key Laboratory of Optical Astronomy and Solar-Terrestrial Environment, School of Space Science and Physics, Institute of Space Sciences, Shandong University, Weihai, Shandong, 264209, China; ruishengzheng@sdu.edu.cn}

\correspondingauthor{Ruisheng Zheng}
\email{ruishengzheng@sdu.edu.cn}


\begin{abstract}
Solar extreme ultraviolet (EUV) waves are large-scale propagating disturbances in the corona. It is generally believed that the vital key for the formation of EUV waves is the rapid expansion of the loops that overlie erupting cores in solar eruptions, such as coronal mass ejections (CMEs) and solar jets. However, the details of the interaction between the erupting cores and overlying loops are not clear, because that the overlying loops are always instantly opened after the energetic eruptions. Here, we present three typical jet-driven EUV waves without CME to study the interaction between the jets and the overlying loops that remained closed during the events. All three jets emanated from magnetic flux cancelation sites in source regions. Interestingly, after the interactions between jets and overlying loops, three EUV waves respectively formed ahead of the top, the near end (close to the jet source), and the far (another) end of the overlying loops. According to the magnetic field distribution of the loops extrapolated from Potential Field Source Surface method, it is confirmed that the birth places of three jet-driven EUV waves were around the weakest magnetic field strength part of the overlying loops. We suggest that the jet-driven EUV waves preferentially occur at the weakest part of the overlying loops, and the location can be subject to the magnetic field intensity around the ends of the loops.

\end{abstract}

\section{Introduction}
Extreme ultraviolet (EUV) waves are spectacular propagating disturbances in the solar corona ~\citep{1998Thompson}. They can provide potential diagnostics on coronal magnetic field strengths and coronal plasma parameters for global coronal seismology~\citep{2013Kwon, 2013long, 2017long}. It is generally believed that EUV waves are best interpreted as the bimodal composition of an outer fast-mode MHD wave and an inner non-wave component of coronal mass ejections (CMEs)~\citep{2011chen, 2014liu, 2013zheng, 2014zheng, 2020mei, 2020zhou, 2021zhou, 2021Chandra}. More details about EUV waves can be found in recent reviews~\citep{2014liu, 2015Warmuth, 2016chen, 2017long}.

EUV waves are associated with various of solar successful eruptions, such as CMEs, solar flares, filament eruptions, and and their launch strongly depends on the rapid lateral expansion of the CME flank~\citep{2013nitta, 2014liu}. In addition, it has been reported that the EUV wave can be generated by the failed eruption or the filament activation without any eruption~\citep{2012zheng, 2018zheng, 2020zheng}. Hence, the formation of EUV waves is subject to the rapid expansion of the loops covering the erupting cores, no matter if associated CMEs happen. However, the interactions between the erupting cores and overlying loops are not clear, because that the overlying loops are usually instantly opened by the intense eruptions.

Comparing with different initial wavefront morphologies, EUV waves are categorized into two groups of flux-rope-driven waves and jet-driven waves, and it is proposed that the formation and morphology of EUV waves depends on the configuration of the overlying loops relative to the erupting cores~\citep{2019zheng}. For flux-rope-driven waves, the erupting flux ropes thrust the overlying loops outward in all directions, and therefore the wave are generated with circular wavefronts ahead of all portions of the overlying loops. For jet-driven waves, the jets emanate from one end of the overlying loops, and the one-way movements result in arc-shaped wavefronts propagating in a limited angular extent. Note that the jet-driven waves were driven by the sudden expansion of coronal loops due to the impingement of the jet upon the loops as what has been pointed out by \cite{2018shena, 2018shenb}, rather than driven by jets directly in a magnetic tube, like the work in~\cite{2018shen}.

In this article, we choose three typical jet-driven waves without any CME to study the interactions between the jets and overlying loops. Three waves all formed ahead of some special parts of the overlying loops (top part, far end part and near end part, respectively) that remained closed during the eruptions. The birth places of the waves reveal the interaction sites between the erupting cores and overlying loops, which can shed light on the generation mechanism of EUV waves.

\section{Observation}
We mainly employed the observations in two different perspectives from Atmospheric Imaging Assembly (AIA;~\cite{2012Lemen}) on board Solar Dynamics Observatory (SDO;~\cite{2012Pesnell}) and from Extreme Ultraviolet Imager (EUVI;~\cite{2008Howard}) onboard the spacecraft A of Solar-Terrestrial Relations Observatory (STEREO). The AIA has 7 EUV wavebands with a cadence of 12 second and a pixel resolution of 0$\farcs$6. The EUVI images have a pixel resolution of 1$\farcs$58, and their cadences are 5 minutes or 10 minutes for 171, 195, and 304~{\AA}. The magnetic field evolution was checked by the Helioseismic and Magnetic Imager (HMI;~\cite{2012Scherrer}) magnetograms, with a cadence of 45 seconds and pixel scale of 0$\farcs$6. The jets and waves were better displayed in AIA composite images. The time-slice approach was employed to analyze the dynamics of the waves and their associated jets. In addition, the coronal magnetic filed lines were extrapolated with the Potential Field Source Surface (PFSS;~\cite{2003Schrijver}) package of SolarSoftWare.

\section{Results}

The three events happened in the active region (AR) 12645 ($\sim$S10W11) on 2017 April 01 (E1), AR 12671 ($\sim$N10E47) on 2017 August 16 (E2), and AR 12498 ($\sim$N19E50) on 2016 February 10 (E3) (the left column in Figure 1). Three source regions all involved a main negative polarity of AR and a nearby parasitic positive polarity (red boxes in the middle column in Figure 1). In the source regions, there existed continuous magnetic interactions for the main polarities and the parasitic polarities (Animation 1). The magnetic flux changes of positive (blue) and negative (red) polarities in the source regions (the right column in Figure 1) clearly show the continuous magnetic flux cancellations and emergences. Remarkably, the main AR polarities all experienced magnetic flux cancellation (red downward arrows) around the onsets jets (dotted vertical lines). It indicates the intimate relationships between the continuous magnetic flux cancellations and the jets. More details about the triggering and formation mechanisms of the corona jet can refer to \cite{2021shen}.

Before the jet onsets, there appeared some brightenings (cyan arrows) above the sites of magnetic cancellations between the main AR polarities (red contours) and parasitic polarities (blue contours) in the source regions (the first column of Figure 2). A few minutes later, the jets ejected (Animation 2 and yellow arrows in the middle column). Note that the jets emanated from one end to another end of the highly overlying loops (blue arrows in panels (c) and (f)). Incidentally, the overlying loops in E3 were faint, but twin dimmings (black arrows in Figure 2(i)) appeared after the jet movements, and their intensities decreased nearly simultaneously. It likely indicates that there existed large-scale overlying loops (white dotted line in panel (i)) connecting twin dimmings in E3. Moreover, the jet in E2 seems like the eruption of a mini-filament towards the loops that in the form of a blowout jet, like the study in \cite{2017shen}. It has difference with the jets in E1 and E3 that the jets were along the loops, but we just focus on the jet in E2 was emanated from one end of the loops to another. Hereafter, the end as the jet sources and another end of overlying loops were separately referred as ``near end" and ``far end''. The near ends and far ends of the overlying loops for three events respectively rooted at the source negative polarities and the remote positive polarities (red and blue contours in the right column).

Following the movements of jets, the overlying loops expanded outwards, which was followed by the loop-like dimmings (white arrows in Figure 3). Sequentially, three EUV waves formed ahead of the expanding loops (Animation 3 and pink dashed lines in the left and middle columns). Intriguingly, the overlying loops remained closed after the loop expansions, and three waves seemed to be born around different parts of the overlying loops. For E1, the wave was primarily generated ahead of the top of the overlying loops (pink dashed lines in panels (a)-(b)). For E2 and E3, the waves mainly formed around the far end and the near of the overlying loops (dashed lines in panels (d) and (g)), respectively, which was also clearly shown by the waves and loop-like dimmings in the edge view of EUVI-A (pink and white arrows in panels (e) and (h)). Moreover, the location relationships between the waves and the overlying loops were confirmed by the wavefront curves (pink dashed lines) and PFSS extrapolated magnetic field lines, superposing on the magnetograms (the right column). It is much more clear that three waves had different birth places along the overlying loops relative to the jet sources.

The wave propagations in some selected slices extending from source regions (yellow dashed lines in Figure 3) are shown in time-distance plots (Figure 4). Two paths were selected in each event that one path is along the wave direction (the wave is clear in this direction), and the other path is along other direction to ensure the wave only generated in the above direction. Specifically, S1a and S1b passed the top and the far end of the overlying loops in E1, S2a and S2b crossed the far end and the top of the overlying loops in E2, and S3a and S3b got through the near end and the top of the overlying loops in E3. In the time-distance plots along S1a-S2a-S3a (the left column), the wave signals were obvious, which separately lasted $\sim$13, 10, and 10 minutes, and their speeds were separately $\sim$524, 346 and 647 km s$^{-1}$. Moreover, the waves had a close temporal-spacial relationships with the associated jets that separately showed speeds of $\sim$360, 190 and 220 km s$^{-1}$. In the time-distance plots along S1b-S2b-S3b (the right column), only the jet in E2 and dimmings due to the loop expansions were clear (red and blue arrows), but it is hardly to distinguish the wave signals. It possibly indicates that three waves predominantly formed ahead of one section of the related overlying loops, consistent with the location relationships between the waves and overlying loops in Figure 3.

To understand three wave formed at different places of the overlying loops, we first focus on the magnetic field intensity surrounding two ends of the overlying loops (Figure 5). For each event, two boxes with a same area enclosed two ends of the extrapolated field lines (yellow boxes in the left column). Since the extrapolated field lines anchor in single polarity patch in the one side in PFSS method, which is also the dominant polarity around the ends of extrapolated field lines, the dominant flux in the box is taken to perform further calculation. In a period of two hours covering each event, the magnetic flux change of the dominant polarities in the boxes was first calculated (red and blue curves in the right column), and then got the average magnetic flux of the dominated polarities in the near end ($\overline{\phi}_{near}$) and far end ($\overline{\phi}_{far}$), and finally obtained the ratio of ($\overline{\phi}_{near}$) to ($\overline{\phi}_{far}$). The ratios for three events were $\sim$0.99, 11.1 and 0.65, respectively. The ratio numbers represent that the magnetic field intensity in the near end was nearly equal to that in the far end for E1, and was much stronger than that in the far end for E2, and was weaker than that in the far end for E3, respectively. Further more, the magnetic field strength ($\mid$B$\mid$) at each point of the extrapolated loops was first calculated through extrapolated three-dimensional field (B$_r$, B$_\theta$, B$_\phi$) in PFSS method, and then superimposed on the loops with different colors (refer to colorbars). For each loop, the weakest point and the weakest part was searched, as represented by the ``X'' symbol and  pink curve, respectively. It is seen that the range of 0\,--\,3 Gauss for the definition of ``weakest part'' well covered the birth place of the wave (pink dashed line) in each event. Accordingly, the weakest part was in the top for E1 and offset to the weak end for E2 and E3, which indicates the good spatial relationship between the birth place of the waves and the weakest part of the loops.

\section{Conclusions and Discussion}
In this article, we present three jet-driven waves without any CME. Before the jet onsets, there existed continuous magnetic flux cancellations in the source regions (Figure 1). Subsequently, there appeared some brightenings in the upper atmosphere, and the jets ejected and moved along the large-scale overlying loops (Figure 2). Because of the interactions between the jets and overlying loops, three jet-driven EUV waves formed with speeds in a range of $\sim$346-647 km s$^{-1}$  and lifetimes of more than ten minutes. As discussed by \cite{2015su} and \cite{2017shen, 2018shen}, jet-driven waves have a relatively shorter lifetime than those associated with CMEs. This is reasonable because the CME can provide a continuous drive to the wave, and the three waves fit this pattern. Intriguingly, the overlying loops remained closed, and the birth places of three waves were predominant around the top, the far end, and the near end of overlying loops, respectively (Figure 3 and 4).

It is very interesting that the birth places of three waves were inconsistent, though three jets emanated from similar near ends. One question arises that what condition determines the birth places of jet-driven waves. The first focus is the magnetic field intensity in the two ends of the overlying loops (Figure 5). Incidentally, we assumed that two ends had similar inclination angles in each event, due to the short distance between two ends. Therefore, the longitudinal magnetic field data using in the ratio measurements of the average magnetic flux can be comparable to that of the 3-dimensional magnetic field. By comparing with the magnetic flux of the dominant polarity around two ends of overlying loops, the birth places of the waves appeared around the top and weak end of the overlying loops. Therefore, we deduced that the wave may form around the weakest magnetic field strength part of the overlying loops. For further, the magnetic field strength along the loops was calculated and superimposed on the loops. It is confirmed that the birth place of the waves formed around the weakest part of the loops.

For EUV waves associated with violent eruptions, it is difficult to study the interactions between the erupting cores and overlying loops, because the impulsive flux ropes or energetic jets always instantly destroyed the overlying loops. That is why we selected these typical jet-driven waves to study the interactions between the jets and overlying loops. Naturally, the jets has to move along the overlying loops (E1 and E3), when their energy was not enough to get rid of the confinements.  However, the jets were searching the weakness of the overlying loops. Finally, the jets successfully found and interacted with the weakest part of the overlying loops, which resulted in the waves ahead of the interaction sites. As a result, the birth places of jet-driven wave were around at the weakest part of the overlying loops. On the other hand, it is also suitable for the ejecting plasma under the loops (E2). They also likely have a preference to search and interact with the weakest part of the overlying loops, since the confinement is weakest in this direction.

In summary, the different birth places of three jet-driven waves in E1-E3 were subject to the magnetic field distribution of the overlying loops. We suggest that the jet-driven waves have a preference to be generated ahead of the weakest part of the overlying loops. Further observations are necessary to verify the results and suggestions.

\begin{acknowledgments}
{\it SDO} is a mission of NASA's Living With a Star Program. The authors thank the teams of {\it SDO} and STEREO for providing the data. This work is supported by grants NSFC 11790303 and 12073016, and the open topic of the Key Laboratory of Solar Activities of Chinese Academy Sciences (KLSA202108).
\end{acknowledgments}

{}

\begin{figure}
\epsscale{1} \plotone{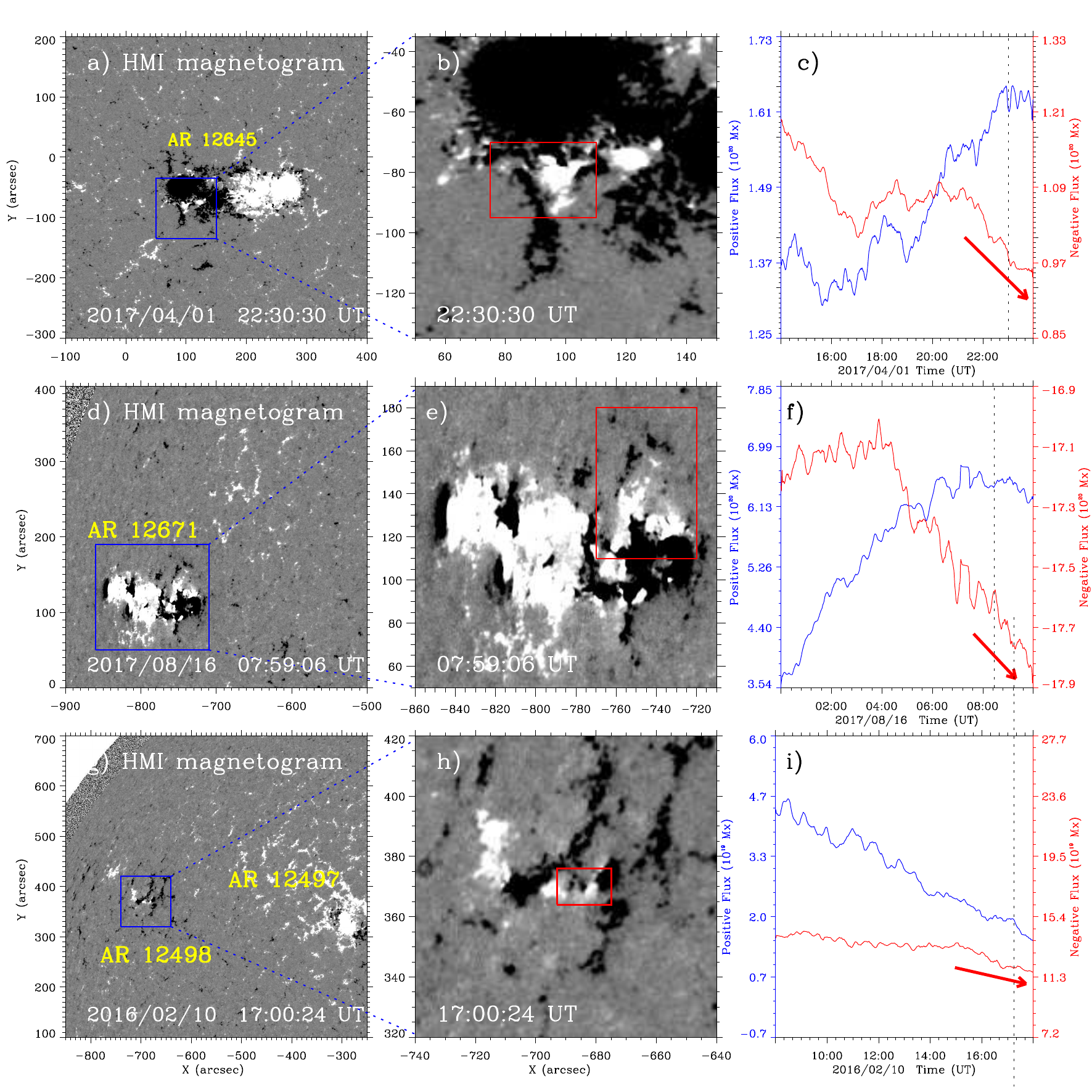}
\caption{The enlarged field of view (FOV) of three source regions in HMI magnetograms (left column). Three source regions in HMI magnetograms (middle column). The blue boxes in left column represent the FOV of source regions in the middle column. The magnetic flux changes of negative AR polarities (red) and positive parasitic polarities (blue) for the red boxes in the middle column (right column). The downward red arrows indicate the decrease of the main AR polarities, and the dotted lines represent the jet onsets. An animation of this figure is available. The animation shows the magnetic field evolution of three source region in HMI magnetograms. In the animation, the periods of three panels are respectively from 13:59:00 to 23:44:45 UT on 2017 April 01, from 23:59:06 to 09:57:36 UT on 2017 Aug 16, and form 07:58:55 to 17:44:39 UT on 2016 Feb 10. The animation does not include the vector plots of c, f and i. (An animation of this figure is available.)
\label{f1}}
\end{figure}

\clearpage
\begin{figure}
\epsscale{0.9} \plotone{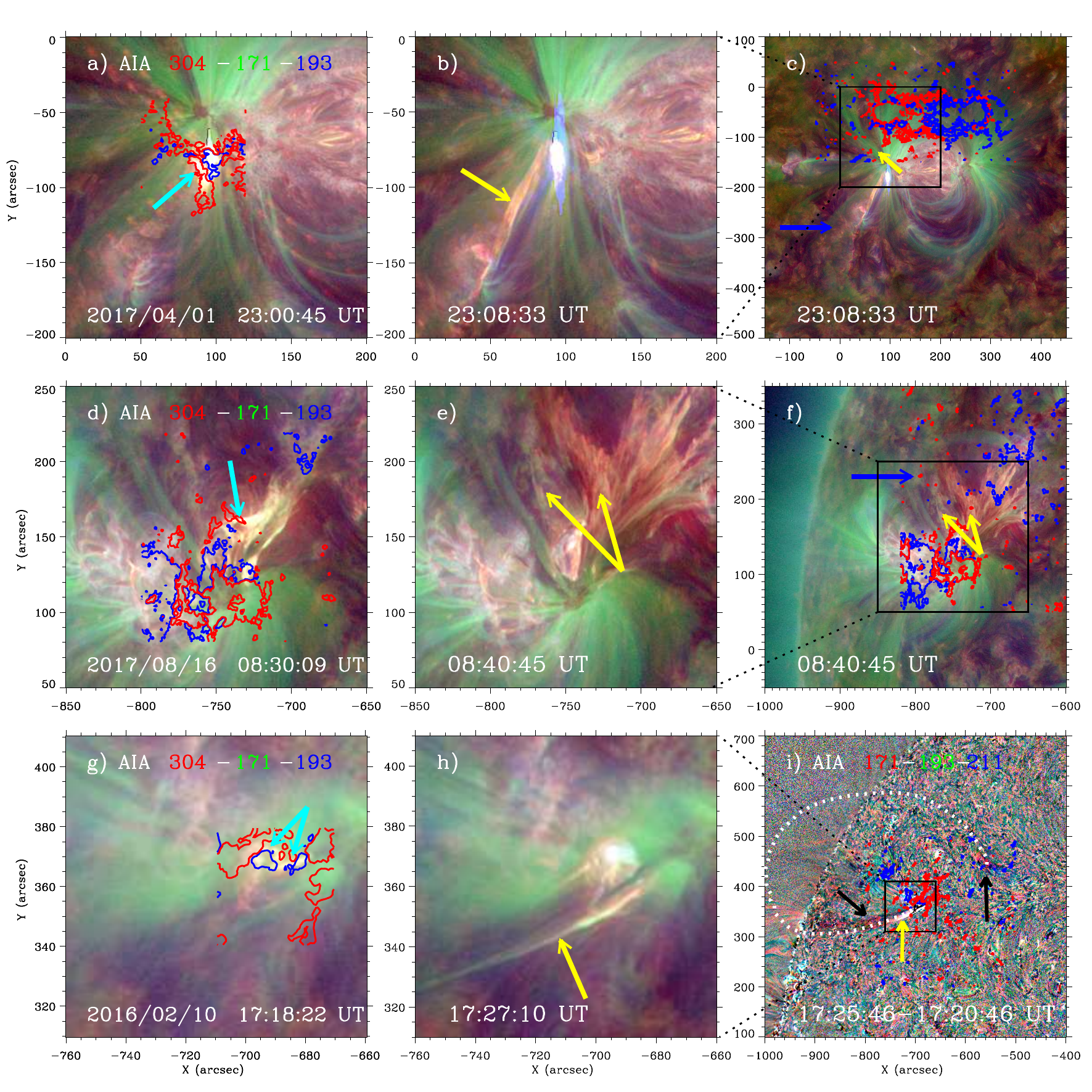}
\caption{Three jets (yellow arrows) in the composite images in AIA 304, 171, and 193~{\AA} (panels (a)-(h)) and in the running-ratio-difference composite image in AIA 171, 193, and 211~{\AA} (panel (i)). The cyan and blue arrows separately indicate the brightenings and the overlying loops. The dotted line and the black arrows show the overlying loops and twin dimmings in E3. The black boxes in the right column represent the FOV of the panels in the left and middle columns. The red and blue contours represent the negative and positive fields in HMI magnetograms. An accompanying animation is available. The animation shows the generation of the jets in composite images. In the animation, the periods of three panels are respectively from 22:50:09 to 23:30:09 UT on 2017 April 01, from 08:10:21 to 08:50:21 UT on 2017 Aug 16, and form 17:00:10 to 17:40:10 UT on 2016 Feb 10. (An animation of this figure is available.)
\label{f2}}
\end{figure}

\clearpage

\begin{figure}
\epsscale{1} \plotone{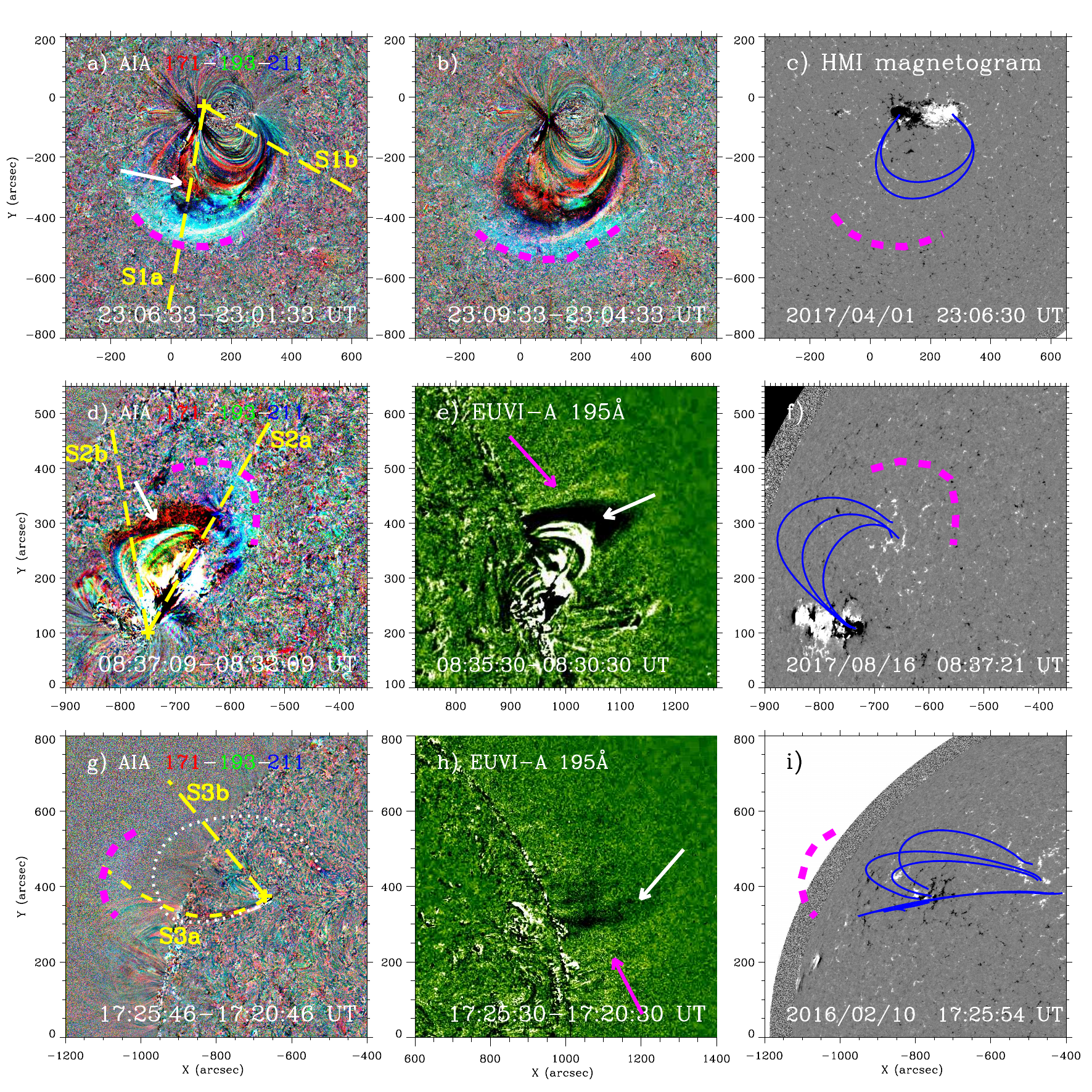}
\caption{Three waves (pink dashed lines and pink arrows) in the composite running-ratio-difference images in AIA 171 (red), 193 (green), and 211~{\AA} (blue) (panels (a), (b), (d), and (g)), and in the running-ratio-difference images in EUVI-A 195~{\AA} (panels (e) and (h)), and in HMI magnetograms superimposed by the PFSS extrapolated magnetic field lines (blue lines in the right column). The yellow dashed lines (S1a, S1b, S2a, S2b, S3a, and S3b) indicate the selected paths to get the time-distance plots in Figure 4. The white arrows show the loop-like dimmings, and the white dotted line represents the overlying loops in E3. An accompanying animation is available. In the animation, the periods of three panels are respectively from 22:59:45 to 23:12:45 UT on 2017 April 01, from 08:30:09 to 08:43:09 UT on 2017 Aug 16, and from 17:19:10 to 17:32:10 UT on 2016 Feb 10. The animation does not include the vector plots of c, f and i. (An animation of this figure is available.)
\label{f3}}
\end{figure}

\clearpage

\begin{figure}
\epsscale{0.8} \plotone{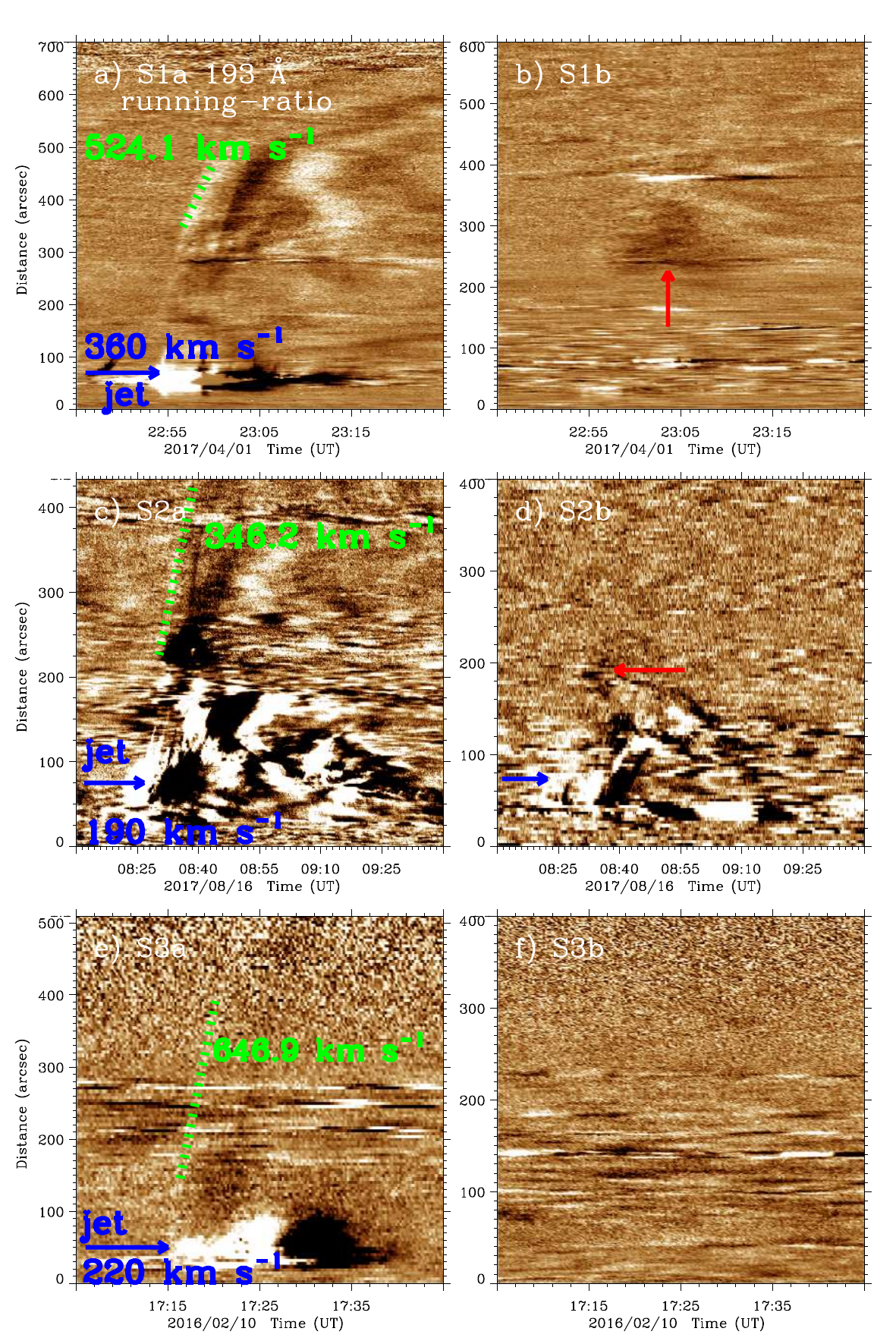}
\caption{The time-distance plots of running-ratio-difference images in AIA 193~{\AA} along the selected paths (yellow dashed lines in the figure 3). The green dotted lines indicate the signals of waves. The blue and red arrows separately indicate the jets and the dimmings. The speeds of the waves and the jets are attached.
\label{f4}}
\end{figure}
\clearpage

\begin{figure}
\epsscale{0.7} \plotone{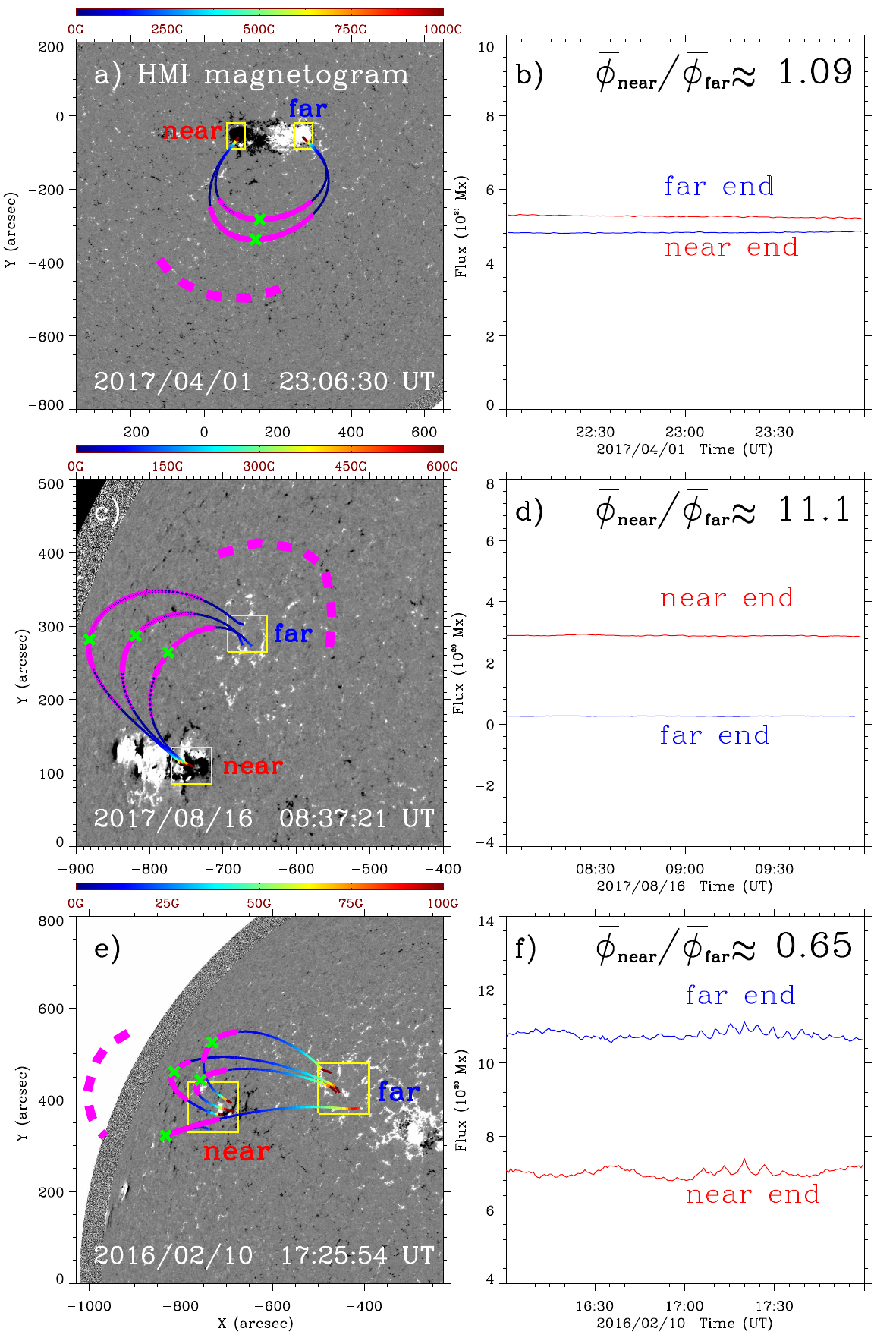}
\caption{The magnetic field distribution of the loops and the comparisons of magnetic field intensities around two ends of the overlying loops. In the left column, the magnetic field strength was superimposed on the loops with different colors (refer to colorbars). The green ``X'' symbols indicate the weakest point of the loops and the pink curves indicate the weakest part of the loops. The yellow boxes with the same area in each panel enclose two ends of the PFSS extrapolated field lines superimposing on HMI magnetograms, and the pink dashed lines represent the waves. In the right column, the red and blue curves indicate the magnetic flux changes of the diminant polarities in the left yellow box. The ratios of the average magnetic flux in the near end to that in the far end are listed.
\label{f5}}
\end{figure}
\clearpage

\end{document}